# R&D on GEM Detectors for Forward Tracking at a Future Electron-Ion Collider

Aiwu Zhang, Vallary Bhopatkar, Marcus Hohlmann, *Member, IEEE*, Xinzhan Bai, Kondo Gnanvo, *Member, IEEE,*
Nilanga K. Liyanage, Matt Posik, Bernd Surrow

*Abstract–*We report the status of R&D on large triple-GEM detectors for a forward tracker (FT) in an experiment at a future Electron Ion Collider (EIC) that will improve our understanding of QCD. We have designed a detector prototype specifically targeted for the EIC-FT, which has a trapezoidal shape with 30.1° opening angle. We are investigating different detector assembly techniques and signal readout technologies, but have designed a common GEM foil to minimize NRE cost for foil production. The assembly techniques comprise either a purely mechanical method including foil stretching as pioneered by CMS but with certain modifications, or gluing foils to frames that are then assembled mechanically, or gluing foils to frames that are then glued together. The first two assembly techniques allow for re-opening chambers so that a GEM foil can be replaced if it is damaged. For readout technologies, we are pursuing a cost-effective one-dimensional readout with wide zigzag strips that maintains reasonable spatial resolution, as well two-dimensional readouts - one with stereo-angle (u-v) strips and another with r-$\phi$ strips. In addition, we aim at an overall low-mass detector design to facilitate good energy resolution for electrons scattered at low momenta. We present design for GEM foils and other detector parts, which we plan to entirely acquire from U.S. companies.

## I. INTRODUCTION

NUCLEAR physics at the frontier of QCD would strongly benefit from experiments at a powerful Electron-Ion Collider (EIC) that is being considered for construction in the USA. These experiments could address the spin problem of hadrons and provide precision measurement of the nucleon structure [1]. Gas Electron Multipliers (GEMs) [2] have been proposed for forward and backward tracking at a future EIC experiment because they have shown very good performance in terms of spatial resolution and long-term stability [3]; they also have been successfully used for tracking in large experiments such as COMPASS [4] and LHCb [5]. We report the status of our R&D activities on prototyping large GEM detectors for an EIC forward (and backward) tracker (FT). This includes the design of a GEM foil of ~1m length, different potential detector assembly techniques, and the design of various signal readout structures.

Manuscript received November 23, 2015. This work was supported by Brookhaven National Laboratory under the EIC eRD6 consortium.

Aiwu Zhang, Vallary Bhopatkar, and Marcus Hohlmann are with Florida Institute of Technology, Melbourne, FL 32901 USA (telephone: 321-674-7275, e-mail: azhang@fit.edu).

Xinzhan Bai, Kondo Gnanvo, and Nilanga K. Liyanage are with University of Virginia, Charlottesville, VA 22904 USA.

Matt Posik and Bernd Surrow are with Temple University, Philadelphia, PA 19122 USA.

## II. THE DETECTOR ASSEMBLY TECHNIQUES

In this section, we explain the three alternative detector assembly techniques that will be investigated by the three groups from Florida Institute of Technology (FIT), University of Virginia (U. Va) and Temple University (TU).

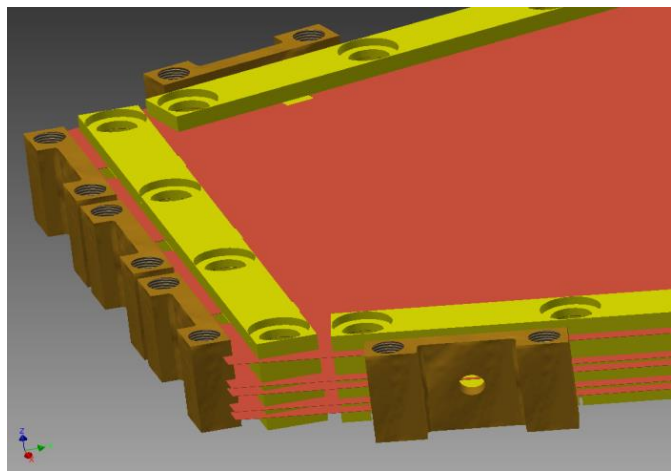

Fig. 1. CAD drawing showing the Triple-GEM detector design to be assembled with the modified mechanical stretching technique. Five foils (red) are sandwiched between small inner frame pieces (olive green) that get stretched against posts (brown). The posts themselves are sandwiched between large stiff outer frames (not shown) to provide mechanical stability.

The FIT group will use a mechanical GEM stretching technique that was pioneered by the CMS GEM collaboration for the CMS muon endcap upgrade [6]. The advantages of this technique are: (1) there are no spacers in the active area and hence no dead area; (2) an assembled detector can be re-opened to swap out a GEM foil in case of damage. The original mechanical stretching technique uses a stack of 3 GEM foils, but puts the drift electrode and readout strips on solid printed circuit boards (PCBs). However, this design results in significant amount of material within the active detector area, which is not optimal for EIC tracking detectors. The potential problems are multiple Coulomb scattering of charged particles and a non-negligible probability for electron showering in the material, which causes systematic effects in the measurement of scattered electron momenta that can be hard to track.

Consequently, we modify the original mechanical stretching technique by designing a larger stack of 5 foils (3

GEM foils, 1 drift foil, and 1 readout foil) that is sandwiched between two outer frames made of stiff material and stretched against posts that are also sandwiched between the outer frames (Fig. 1). Both frames will have thin windows, e.g. made from 100 μm aluminized mylar foil, so that the material in the active area is minimized. High-strength but low-mass material such as carbon fiber will be investigated for the frames since PCB frames may not be strong enough to hold their shapes under the stress required to stretch the foils. An outer frame surrounding all posts will seal the gas volume.

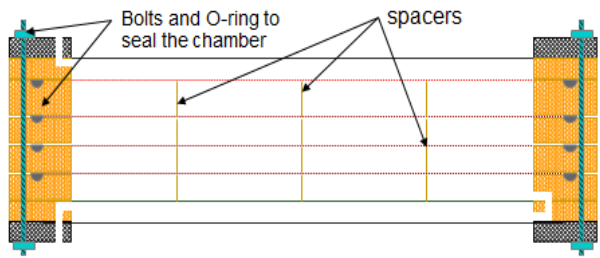

Fig. 2. A schematic that shows the assembly technique with spacer frames.

The U. Va. group plans to glue GEM foils, drift and readout foils to spacer frames, which are held together with screws, and then close the detector with bolts and an O-ring (Fig. 2). In this technique, there will be spacers in the GEM active area, and a frame/foil assembly can also be swapped out in case of damage. Light material will be used for spacer frames that support the GEMs, while high-strength materials such as carbon fiber or ceramic will be used for the external supporting frames.

The TU group will glue GEM foils, drift and readout foils to frames and glue all frames together. Such a detector cannot be opened anymore and foils cannot be changed; however, the overall detector material will be minimized.

## III. THE COMMON GEM FOIL DESIGN

We have designed a common EIC FT GEM foil (Fig. 3) which simultaneously satisfies all the requirements for the three different assembly techniques. The foil will have a trapezoidal shape with a 30.1° opening angle of the active area, so that a full disk can be assembled from 12 such detectors in the endcap region at a collider experiment with some overlap between adjacent detectors for complete coverage. This GEM prototype is intended to provide acceptance very close to the beam pipe, so we choose 80 mm as the distance from the inner edge of the active area at small radius to the vertex of the trapezoid that coincides with the beam. The width of the active area at smallest radius is 80 mm × tan(15.05°) × 2 = 43 mm. Since the width of the raw foil material is limited to 610 mm and a 25 mm margin is required around the foil perimeter for mounting it during GEM production, we design the maximum width of the usable area at the large end to be 560 mm (including 15 mm of space allocated to be sandwiched between the GEM stack frame pieces). The active area of this common GEM foil is 2584 cm². The top surface of the GEM is divided into 8 HV sectors along the R direction at inner radius and into 16 HV sectors along the azimuthal direction at outer radius with ~107 cm² for each HV sector, so that the energy of potential discharges is minimized.

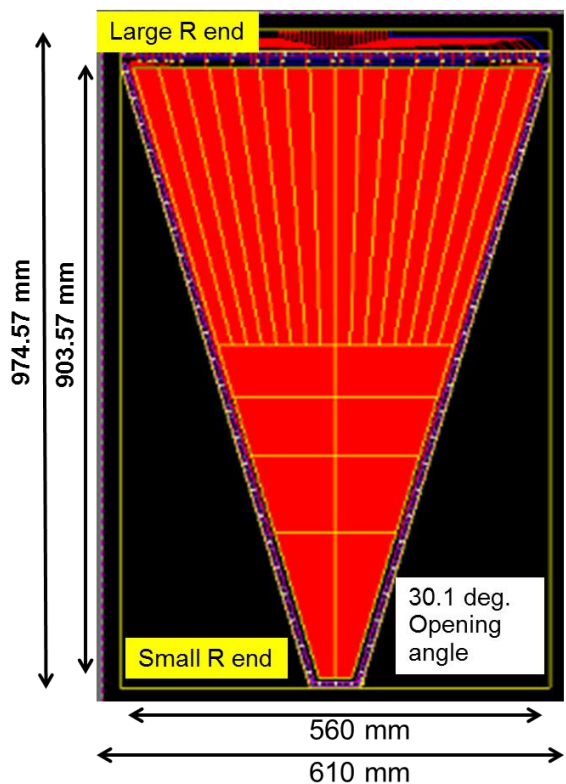

Fig. 3. Sketch of the common EIC FT GEM foil designed with Altium.

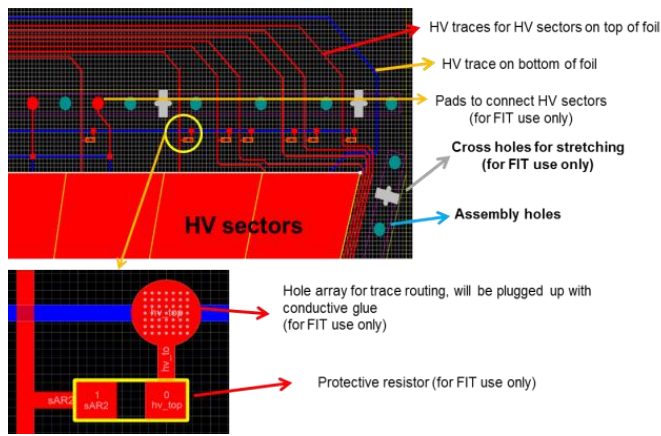

Fig. 4. Zoomed view of the top right corner of the common foil design in Fig. 3 highlighting features required by the different assembly techniques.

Fig. 4 shows some details of how the required features for the three assembly techniques are implemented in the common foil. There are 124 foil-stack assembly holes (4 mm diameter) around the active area; they are distributed with a pitch of 22 mm at the wide end and 20 mm in the other regions. Sixty additional cross-shaped holes allow for screws to be imbedded into the foil stack for assembly with the mechanical stretching technique.

The different detector assembly techniques require different ways to make HV connections to the HV sectors of the GEM.

For the mechanical stretching technique, we place protective resistors onto the foil for each HV sector and all sectors are connected to a single HV pad through these resistors. These HV pads, 6 on the top surface and 6 on the bottom surface, are put in between the assembly holes. For the other two assembly techniques, each HV sector is to be accessed individually when a detector is closed, so HV traces have to be brought from all sectors to pads. When arranging the HV traces, we first route HV traces from HV sectors to the pads outside on the top surface. The HV traces connecting protective resistors and the HV pads between the assembly holes run partially on the bottom surface and must be brought to the top surface at some points. Since metallized vias are hard to produce on foils, the solution is to put an array of holes (~1.3 mm) inside an HV pad and to put conductive glue (EPO-TEK) in the holes to ensure top-bottom connectivity (Fig. 4).

## IV. THE DIFFERENT READOUT STRUCTURE DESIGNS

In this section we present complementary approaches for reading out the new EIC FT GEM prototypes.

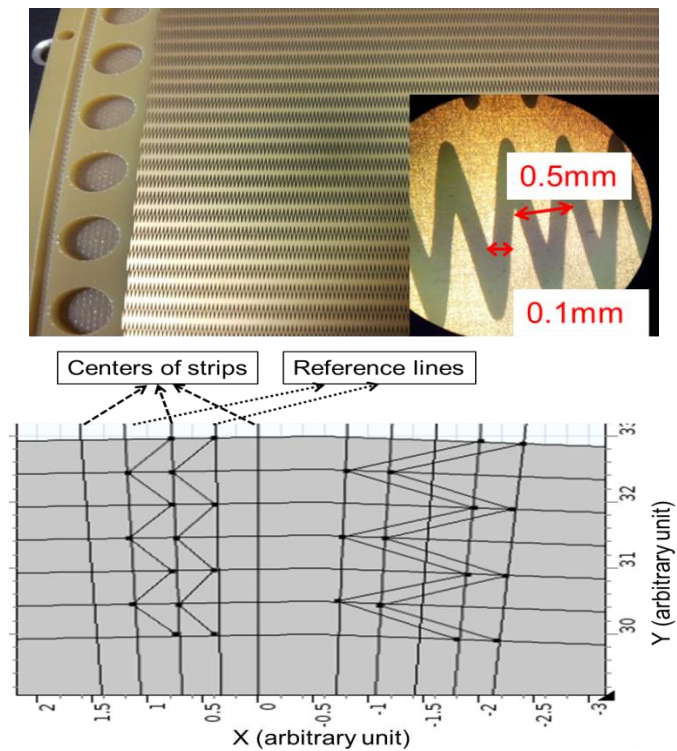

Fig. 5. Top: the zigzag structure for reading out the previous EIC FT GEM prototype [8]. Bottom: two types of zigzag strips running radially; the zigzag strip on the left shows that tips on one strip do not exceed two reference lines, while on the right the tips on a zigzag strip reach the centers of its neighbor strips.

The FIT group has been investigating zigzag strips for reading out GEM detectors [7,8]. A zigzag strip can be designed in such a way that it occupies a bigger area than a normal straight strip with a few hundred micrometers width does, so that using zigzag strips can reduce the number of channels for a GEM detector. The "zigs" and "zags" on a zigzag strip interleave with those of a neighbor strip so that charge sharing between strips can be enhanced and good spatial resolution is preserved. This will be very cost effective for large-area GEM detectors since the number of electronic channels can be significantly reduced.

Fig. 5 (top) shows the zigzag strips for reading out our previous 1-m GEM prototype where the zigzag strips run radially and measure azimuthal φ coordinates of hit positions. In that design, the locations of the tips on a zigzag strip were constrained by only two reference lines (Fig. 5, bottom left) with limited interleaving of zigs and zags and a central "spine" on each strip. This resulted in a nonlinear response when constructing hit positions from the charge centroid method [8]. In the next EIC FT GEM prototype design, each zigzag strip will reach the center lines of its two neighbor strips (Fig. 5, bottom, on the right) to optimize charge sharing so that the response is more linear.

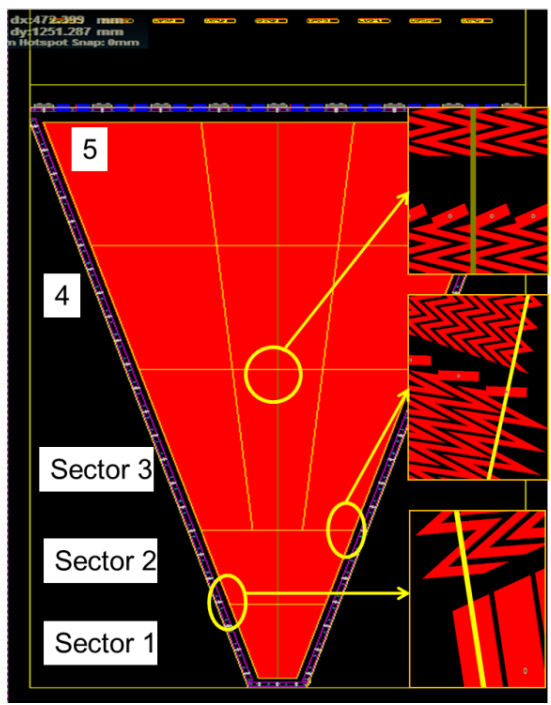

Fig. 6. The readout design (with both straight strips and zigzag strips) for the EIC FT GEM prototype with five different sectors. Design details are show in zoomed-in views of three different regions on the right.

Fig. 6 shows an overview of the full zigzag readout design for the EIC FT GEM prototype. We divide the readout area into five regions. In sector 1 at smallest radius, which extends 12 cm in radius, we use 128 straight strips in order to achieve best spatial resolution and low occupancy since the sector width is small; here the angular pitch of the strips is 4.14 mrad. Sector 2 also has a radial length of 12 cm, but is read out with 128 zigzag strips, so it also has 4.14 mrad angular pitch. Sectors 3-5 have a radial length of ~22 cm each and 384 zigzag strips, so the angular pitch in these sectors is 1.37 mrad. Nine APV hybrids (1152 channels) [9] placed at the wide end of the trapezoid suffice to read out the full detector area. This is the most economical readout design under consideration. In order to reduce detector material, we plan to produce the readout strips either on a GEM-like foil (50 μm thickness) or on flexible PCB material (< 200 μm). Routing strips to plug-in connectors for APV hybrids requires vias or via-like structures.

The U. Va. group will be reading out the GEM detector with two-dimensional stereo-angle strips (u-v strips) [10]. Fig. 7 shows the corresponding design of a readout foil for the next prototype. The strips within each direction are parallel to each other with a pitch of 400 μm; the angle between u and v directions is 30 degrees. This design requires 2560 strips and 20 APV hybrids placed at the wide end of the trapezoid to read out the full detector.

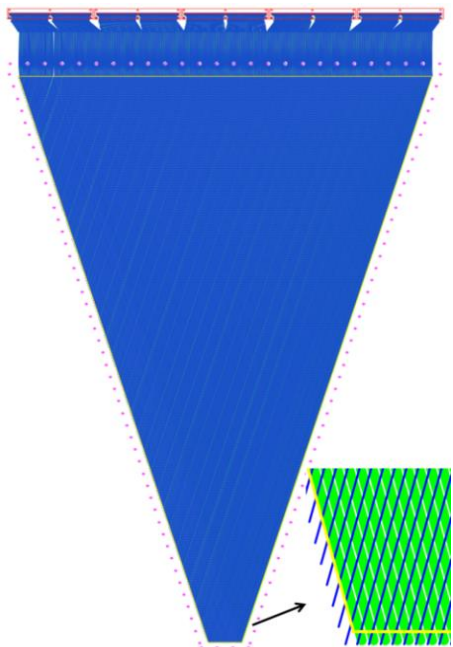

Fig. 7. The two-dimensional long stereo-angle strips (u-v strips) for the EIC FT GEM prototype.

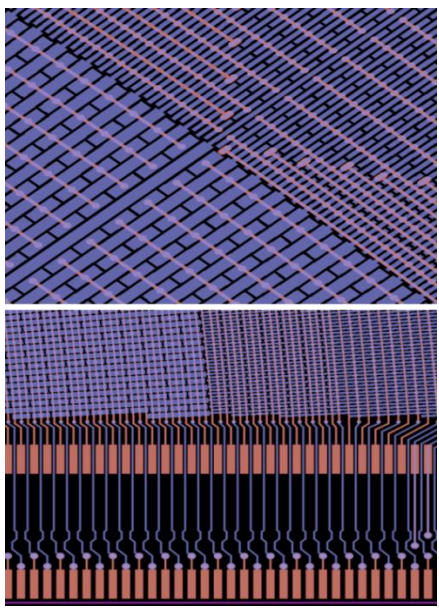

Fig. 8. The two-dimensional strip-pad readout design for R and φ coordinates [11].

The Temple U. group will investigate a two-dimensional readout of R and φ coordinates using a strip-pad design which was demonstrated in the Forward GEM Tracker for the STAR experiment [11] (Fig. 8). Vias are used to make the connections from pad to pad and from strips to connectors for APV hybrids.

## V. SUMMARY AND OUTLOOK

We have designed a 1-m long common GEM foil for forward tracker prototypes for a future EIC detector. The design has been transferred to CERN and we are basically ready to start producing these foils. We will use different techniques to assemble the GEM prototypes; low-mass but stiff materials will be used as much as possible in the detector construction. Complementary readout technologies will be tested; the designs of these readout structures are nearing completion. We expect to begin construction of the new GEM prototypes in 2016.

## REFERENCES

[1] The Electron-Ion Collider White Paper, arXiv:1212.1701.
[2] F. Sauli, Nuclear Instruments and Methods in Physics Research Section A 386 (1997) 531-534.
[3] E. C. Aschenauer et al., eRHIC Design Study: An Electron-Ion Collider at BNL, arXiv: 1409.1633.
[4] B. Ketzer et al., Nuclear Instruments and Methods in Physics Research Section A 535 (2004) 314-318.
[5] A. Cardini, G. Bencivenni, and P. De Simone, "The Operational Experience of the Triple-GEM Detectors of the LHCb Muon System: Summary of 2 Years of Data Taking," 2012 IEEE Nuclear Science Symposium and Medical Imaging Conference, p. 759-762, Oct. 27-Nov. 3, 2012, doi: 10.1109/NSSMIC.2012.6551204.
[6] CMS Technical Design Report for the Muon Endcap GEM Upgrade, CERN-LHCC-2015-012, CMS-TDR-013, ISBN 978-92-9083-396-2 (2015).
[7] D. Abbaneo et al., "Beam Test Results of New Full-Scale Prototypes for CMS High-Eta Muon System Future Upgrade," 2012 IEEE Nuclear Science Symposium and Medical Imaging Conference, p. 1172-1176, Oct. 27-Nov. 3, 2012, doi: 10.1109/NSSMIC.2012.6551293.
[8] A. Zhang et al., "Performance of a Large-area GEM Detector Read Out with Wide Radial Zigzag Strips," submitted to Nuclear Instruments and Methods in Physics Research Section A for publication.
[9] L. Jones, APV25-S1 User Guide, Version 2.2, available at https://cds.cern.ch/record/1069892/files/cer-002725643.pdf.
[10] K. Gnanvo et al., "Performance in Test Beam of a Large-area and Light-weight GEM detector with 2D Stereo-Angle (U-V) Strip Readout," accepted for publication in Nuclear Instruments and Methods in Physics Research Section A, doi: 10.1016/j.nima.2015.11.071.
[11] B. Surrow, "The STAR Forward GEM Tracker," Nuclear Instruments and Methods in Physics Research Section A, 617 (2010) 196-198.